\documentclass[preprint,aps,showpacs]{revtex4}
\usepackage{graphicx,epsf}
\usepackage{lscape}
\usepackage{citesort}
\usepackage[colorlinks,citecolor=blue,urlcolor=red,linkcolor=red]{hyperref}
\begin{document}

\title{{\Large{\bf  Coupling constants of  bottom (charmed)
mesons with pion from
three point QCD sum rules}}}

\author{\small
\small  M.
Janbazi$^1$ \footnote {e-mail: mehdijanbazi@yahoo.com}, N. Ghahramany$^2$ \footnote{e-mail: ghahramany @ susc.ac.ir}, E. Pourjafarabadi$^1$ \footnote{e-mail: pesmaiel@yahoo.com}}
\affiliation{$^1$Department of Physics, Shiraz Branch Islamic Azad University, Shiraz, Iran\\$^2$Physics Department, Shiraz University, Shiraz
71454, Iran}

\begin{abstract}
In this article, the three point QCD sum rules is used to compute the strong coupling constants of vertices containing the strange bottomed ( charmed ) mesons with pion. The coupling constants are calculated, when both the bottom ( charm ) and pion states are off-shell. A comparison of the obtained results of coupling constants with the existing predictions is also made. \par

\end{abstract}


\maketitle
\section{Introduction}
During last ten years, there have been numerous published research articles devoted to the precise determination of the strong form factors and coupling constants of meson vertices via QCD sum rules (QCDSR) \cite{sumrules}. QCDSR formalism have also been successfully used to study some of the " exotic " mesons made of quark- gluon hybrid ($ q\bar{q}g $), tetraquark states ($ q\bar{q}q\bar{q} $), molecular states of two ordinary mesons, glueballs and many others \cite{exotic}. Coupling constants can provide a real possibility for studying the nature of the bottomed and charmed pseudoscalar and axial vector mesons. More accurate determination of these coupling constants play an important role in understanding of the final states interactions in the hadronic decays of the heavy mesons. Our knowledge of the form factors in hadronic vertices is of crucial importance to estimate hadronic amplitudes when hadronic degrees of freedom are used. When all of the particles in a hadronic vertex are on mass-shell, the effective fields of the hadrons describe point-like physics. However, when at least one of the particles in the vertex is off-shell, the finite size effects of the hadrons become important. The following coupling constants have been determined by different research groups:
 $D^* D \pi$
\cite{FSNavarra,MNielsen}, $D D \rho$\cite{MChiapparini}, $D^* D
\rho$\cite{Rodrigues3}, $D^* D^*
\rho$\cite{MEBracco}, $D D J/\psi$ \cite{RDMatheus},  $D^* D J/\psi$\cite{RRdaSilva},
 $D^* D^* J/\psi$ \cite{EBracco}, $D_s D^* K$,
$D_s^* D K$ \cite{ALozea}, $D D \omega$  \cite{LBHolanda}
and $VD_{s0}^{*} D_{s0}^{*}$, $VD_{s}
D_{s}$, $VD_{s}^{*} D_{s}^{*}$ and
$VD_{s1} D_{s1} $\cite{Janbazi}, in the framework of three point QCD sum rules. It is very
important to know the precise functional form of the form factors in
these vertices and even to know how this form changes when one or
the other (or both) mesons are off-shell \cite{Janbazi}.

In this review, we focus on the method
of three point QCD sum rules to calculate, the strong
form factors and coupling constants associated with the $B_1B^*\pi$, $B_1B_0\pi$, $B_1B_1\pi$, $D_1D^*\pi$, $D_1D_0\pi$ and $D_1D_1\pi$ vertices, for both the bottom (charm) and pion
states being off-shell.
The three point correlation function is investigated in two
phenomenological and theoretical sides.
In physical or phenomenological part, the
representation is in terms of hadronic
degrees of freedom which is
responsible for the introduction of the form
factors, decay constants and masses.
In QCD or theoretical part, which consists of two, perturbative
and non-perturbative contributions (In the present work
the calculations contributing the quark-quark and
quark-gluon condensate diagrams are considered as non-perturbative
effects), we
evaluate the correlation function in quark-gluon language and in
terms of QCD degrees of freedom such as, quark
condensate, gluon
condensate, etc, by the help of the Wilson
operator product
expansion(OPE). Equating two sides and
applying the double Borel
transformations, with respect to the momentum
of the initial and final states, to suppress
the contribution of the higher states and
continuum, the strong form factors are estimated.

The outline of the paper is as follows.
In section II, by introducing the sufficient correlation
functions, we obtain QCD sum rules for the
strong coupling constant of the considered
$B_1B^*\pi$, $B_1B_0\pi$ and $B_1B_1\pi$
vertices. With the necessary changes in quarks, we
can easily apply the same calculations to the
$D_1D^*\pi$, $D_1D_0\pi$ and $D_1D_1\pi$ vertices .
In obtaining the sum rules for physical
quantities, both light quark-quark
and light quark-gluon condensate diagrams are considered as
non-perturbative contributions. In section III,
the obtained sum rules
for the considered strong coupling constants are numerically analysed.
We will obtain the numerical values for each
coupling constant when both the bottom (charm) and pion
states are off-shell. Then taking the average of the
two off-shell cases, we will obtain final numerical
values for each coupling constant. In this section,
we also compare our results with the existing
predictions of the other works.
\section{ THE THREE POINT QCD SUM RULES METHOD}
\begin{figure}[th]
\begin{center}
\begin{picture}(130,23)
\put(-17,-15){ \epsfxsize=15cm \epsfbox{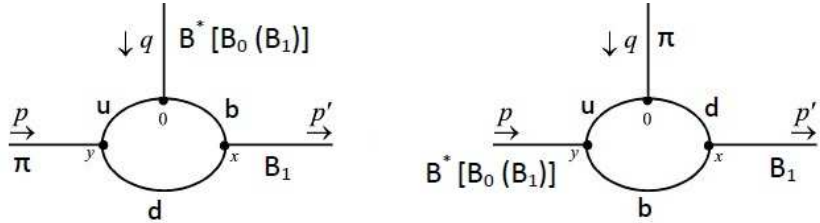} }
\end{picture}
\end{center}
\vspace*{-0.5cm} \caption{perturbative diagrams for off-shell bottom (left)
and off-shell pion (right).}\label{F1}
\end{figure}

 In order to evaluate the strong coupling constants, it is necessary to know the effective Lagrangians of the interaction which, for the vertices $ B_{1}B^*\pi$, $ B_{1}B_{0}\pi$ and $ B_{1}B_{1}\pi$, are\cite{Song12,123}:
\begin{eqnarray}
{\cal L}_{B_1B^*\pi}&=&g_{B_1B^*\pi} B_{1}^{\alpha}(\pi^{+} B^{*-}_{\alpha}- \pi^{-} B^{*+}_{\alpha}), \nonumber \\
{\cal L}_{B_1B_0\pi}&=&ig_{B_1B_0\pi} B_1^{\alpha}( B_0^{-} \partial_{\alpha} \pi^{+} -\partial_{\alpha}B_0^{-} \pi^{+} )+H.c. , \nonumber \\
{\cal L}_{B_1B_1\pi}&=&-g_{B_1B_1\pi} \epsilon^{
\alpha\beta \gamma \sigma}\partial_{\alpha} B_{1\beta}(\partial_{\gamma} B_{1\sigma}^{+} \pi^{-} +\pi^{+} \partial_{\gamma} B^{-}_{1\sigma} ), 
\end{eqnarray}
From these Lagrangians, we can extract elements associated with the $ B_{1}B^*\pi$, $ B_{1}B_{0}\pi$ and $ B_{1}B_{1}\pi$  momentum dependent vertices, that can
be written in terms of the form factors:
\begin{eqnarray}\label{eq21}
\langle  B_{1}(p', \epsilon') |  B^*(p,\epsilon) \pi(q) \rangle &=& g_{ B_{1}B^*\pi}(q^2)  (\epsilon'.\epsilon) \frac{p.q}{m_{B1}},\nonumber \\
\langle B_{1}(p', \epsilon') | B_{0}(p)  \pi(q) \rangle &=&
g_{B_{1}B_{0}\pi}(q^2)   \epsilon'.q,\nonumber\\
\langle  B_1(p', \epsilon') |  B_1(p,\epsilon) \pi(q) \rangle &=&i g_{ B_1B_1\pi}(q^2) \epsilon^{
\alpha\beta \gamma \sigma}
 \epsilon'_\gamma(p') \epsilon_\sigma(p) p'_\beta q_\alpha,
\end{eqnarray}
where $p$ and $p'$ are the four
momentum of the initial and final mesons and  $q=p'-p$,  $\epsilon$
and $\epsilon'$ are the polarization vector of the $B^*$  and
$B_{1}$ mesons. We study the strong coupling constants $ B_{1}B^*\pi$, $ B_{1}B_{0}\pi$ and $ B_{1}B_{1}\pi$ vertices
when both $\pi$ and $B^*[B_{0}(B_1)]$ can be off-shell.
The interpolating currents $j^{\pi}=\bar q \gamma_5 q$, $j^{B_{0}}=\bar{q} Q$,
$j_{\nu}^{B^*}=\bar{q} \gamma_{\nu} Q$ and
$j_{\mu}^{B_{1}}=\bar{q} \gamma_{\mu} \gamma_5 Q$ are interpolating
currents of $\pi$, $B_{0}$, $B^*$, $B_{1}$ mesons, respectively with $q$ being the up or down and $Q$ being the heavy quark fields. We write the three-point correlation function associated with the  $ B_{1}B^*\pi$, $ B_{1}B_{0}\pi$ and $ B_{1}B_{1}\pi$ vertices. For the off-shell $B^*[B_{0}(B_1)]$ meson, Fig.\ref{F1} (left), these correlation functions are given
by:
\begin{eqnarray}\label{eq22}
\Pi^{B^*}_{\mu\nu}(p, p')=i^2 \int d^4x d^4y
e^{i(p'x-py)}\langle 0 |\mathcal{T}\left\{j_{\mu}^{B_{1}}(x)
{j_{\nu}^{B^*}}^{\dagger}(0) {j^{\pi}}^{\dagger}(y)\right\}| 0 \rangle,
\end{eqnarray}
\begin{eqnarray}\label{eq23}
\Pi^{B_0}_{\mu}(p, p')=i^2 \int d^4x d^4y
e^{i(p'x-py)}\langle 0 |\mathcal{T}\left\{j_{\mu}^{B_{1}}(x)
{j^{B_0}}^{\dagger}(0) {j^{\pi}}^{\dagger}(y)\right\}| 0 \rangle,
\end{eqnarray}
\begin{eqnarray}\label{eq24}
\Pi^{B_1}_{\mu\nu}(p, p')=i^2 \int d^4x d^4y
e^{i(p'x-py)}\langle 0 |\mathcal{T}\left\{j_{\mu}^{B_{1}}(x)
{j_{\nu}^{B_1}}^{\dagger}(0) {j^{\pi}}^{\dagger}(y)\right\}| 0 \rangle,
\end{eqnarray}
and for the off-shell $\pi$ meson, Fig.\ref{F1} (right), these quantities
are:
\begin{eqnarray}\label{eq25}
\Pi^{\pi}_{\mu\nu}(p, p')=i^2 \int d^4x d^4y
e^{i(p'x-py)}\langle 0 |\mathcal{T}\left\{j_{\mu}^{B_{1}}(x)
{j^{\pi}}^{\dagger}(0) {j_{\nu}^{B^*}}^{\dagger}(y)\right\}| 0 \rangle,
\end{eqnarray}
\begin{eqnarray}\label{eq26}
\Pi^{\pi}_{\mu}(p, p')=i^2 \int d^4x d^4y
e^{i(p'x-py)}\langle 0 |\mathcal{T}\left\{j_{\mu}^{B_{1}}(x)
{j^{\pi}}^{\dagger}(0) {j^{B_0}}^{\dagger}(y)\right\}| 0 \rangle,
\end{eqnarray}
\begin{eqnarray}\label{eq27}
\Pi^{\pi}_{\mu\nu}(p, p')=i^2 \int d^4x d^4y
e^{i(p'x-py)}\langle 0 |\mathcal{T}\left\{j_{\mu}^{B_{1}}(x)
{j^{\pi}}^{\dagger}(0) {j_{\nu}^{B_1}}^{\dagger}(y)\right\}| 0 \rangle,
\end{eqnarray}

Correlation function  in (Eqs. (\ref{eq22} - \ref{eq27}))
in the OPE and in the phenomenological side can be written in terms of several tensor
structures. We can write a sum rule to find the coefficients of each structure, leading to 
as many sum rules as structures. In principle all the structures should yield the
same final results but, the truncation of the OPE changes different structures in different ways. Therefore some structures lead to sum rules which are
more stable. In the simplest cases, such as in the $B_1B^*\pi$ vertex, we have five structures $g_{\mu\nu}$, $p_{\mu}p_{\nu}$, $p_{\mu}p'_{\nu}$, $p'_{\mu}p_{\nu}$ and $p'_{\mu}p'_{\nu}$  . We have selected the $g_{\mu\nu}$ structure. In this structure the quark condensate (the
condensate of lower dimension) contributes in the case of bottom meson off-shell. We also did the
calculations for the structure $p'_{\mu}p_{\nu}$ and the final results of both structures in predicting of
$g_{\mu\nu}$  are the same for $g_{B_1B^*\pi}$ and in the $B_1B_0\pi$ vertex, we have two structure $p'_{\mu}$ and $p_{\mu}$. The two structures give the same
result for $g_{B_1B_0\pi}$. We have chosen the $p'_{\mu}$ structure. In the $B_1B_1\pi$ vertex we have only one structure $\epsilon^{ \alpha\beta \mu \nu} p_\alpha p'_\beta$ is written as:
\begin{eqnarray}\label{eq28}
\Pi^{B^*(\pi)}_{\mu\nu}(p^2, p'^2,q^2)&=&(\Pi^{B^*(\pi)}_{per}+\Pi^{B^*(\pi)}_{nonper})
g_{\mu\nu}+\cdots,\nonumber \\
\Pi^{B_0(\pi)}_{\mu}(p^2, p'^2,q^2) &=& (\Pi^{B_0(\pi)}_{per}
+\Pi^{B_0(\pi)}_{nonper})p'_{\mu}+\cdots,\nonumber \\
\Pi^{B_1(\pi)}_{\mu\nu}(p^2, p'^2,q^2)&=&(\Pi^{B_1(\pi)}_{per}+\Pi^{B_1(\pi)}_{nonper})
\epsilon^{ \alpha\beta \mu \nu} p_\alpha p'_\beta
\end{eqnarray}
where $\cdots$ denotes other structures and higher states.

The phenomenological side of the vertex function is obtained
by considering the contribution of three complete sets of
intermediate states with the same quantum number that should
be inserted in Eqs.(\ref{eq22} - \ref{eq27}).
We use the standard definitions for the decay constants
$f_M$ ($ f_{\pi} $ , $ f_{B_0} $ , $ f_{B^*} $ and $ f_{B_{1}} $) and are given by:
\begin{eqnarray}\label{eq29}
\langle 0 | j^{\pi} | \pi(p) \rangle &=& \frac{m_{\pi}^2 f_{\pi}}{m_u+m_d}, \nonumber \\
\langle 0 | j^{B_{0}} | B_{0}(p) \rangle &=& m_{B_{0}} f_{B_{0}},\nonumber \\
\langle 0 | j_{\nu}^{B^*} | B^*(p, \epsilon) \rangle &=& m_{B^*}
f_{B^*} \epsilon_{\nu}(p), \nonumber \\
\langle 0 | j_{\mu}^{B_{1}} | {B_{1}}(p', \epsilon') \rangle &=&
m_{B_{1}} f_{B_{1}} \epsilon'_{\mu}(p'),
\end{eqnarray}
 The phenomenological part for the $g_{\mu\nu}$ structure associated to $B_1B^*\pi $ vertex, when
$B^*(\pi)$ is off-shell  meson is as follow:
\begin{eqnarray}\label{eq210}
\Pi^{B^*(\pi)}_{\mu\nu}&=&-g_{B_{1}B^*\pi}^{B^*(\pi)}(q^2)
\frac{ m_{\pi}^2 m_{B^*}f_{\pi} f_{B^*} f_{B_1}(m_{B_{1}}^2-
m_{\pi(B^*)}^2-q^2)}{2(q^2-m_{B^*(\pi)}^2)(p^2-m_{\pi(B^*)}^2)
(p'^2-m_{B_{1}}^2)(m_u+m_d)}g_{\mu\nu} +h.r,
\end{eqnarray}
The phenomenological part for the $p'_{\mu} $ structure
 related to the $B_1B_0\pi $ vertex, when $B_0(\pi)$ is off-shell  meson is:
 \begin{eqnarray}\label{eq211}
\Pi^{B_0(\pi)}_{\mu}&=&-g_{B_{1}B_0\pi}^{B_0(\pi)}(q^2)
\frac{ m_{\pi}^2m_{B_0}m_{B^*} f_{\pi} f_{B^*} f_{B_{1}}(m_{B_{1}}^2+
m_{\pi(B_0)}^2-q^2)}{2(q^2-m_{B_0(\pi)}^2)(p^2-m_{\pi(B_0)}^2)
(p'^2-m_{B_{1}}^2)(m_u+m_d)} p'_{\mu}+h.r,
\end{eqnarray}
The phenomenological part for the $\epsilon^{ \alpha\beta \mu \nu} p_\alpha p'_\beta$ structure
 related to the $B_1B_1\pi $ vertex, when $B_1(\pi)$ is off-shell  meson is:
\begin{eqnarray}\label{eq212}
\Pi^{B_1(\pi)}_{\mu\nu}&=&-i g_{B_1B_1\pi}
^{B_1(\pi)}(q^2)\frac{m_{\pi}^2 m_{B_1} ^2   f_{\pi} f_{B_1}^2 }{(q^2-m_{B_1(\pi)}^2)
(p^2-m_{\pi(B_1)}^2)(p'^2-m_{B_1}^2)(m_u+m_d)}\epsilon^{ \alpha\beta \mu \nu} p_\alpha p'_{\beta}+h.r,\nonumber\\
\end{eqnarray}
In the Eqs.(\ref{eq210} - \ref{eq212}), h.r. represents the
contributions of the higher states and continuum.

With the help of the operator product expansion (OPE) in Euclidean
region, where $p^2,p'^2\to -\infty$, we calculate the QCD side of
the correlation function (Eqs. (\ref{eq22} - \ref{eq27}))
containing perturbative and non-perturbative parts.
 In practice, only the first few condensates contribute significantly, the
most important ones being the 3-dimension, $\langle\bar{d}d\rangle$, and the 5-dimension, $\langle\bar{d}\sigma_{\alpha \beta}
T^{a}G^{a\alpha\beta}d\rangle$, condensates.
For each invariant structure, i, we can write
\begin{eqnarray}\label{eq213}
\Pi^{(theor)}_{i}(p^2, p'^2, q^2) &=& -\frac{1}{4 \pi^2} \int_{(m_d+m_b)^2}^{\infty}
ds'\int_{s_{1(2)}}^{\infty} ds\frac{\rho_{i}(s, s',
q^2)}{(s-p^2)(s'-p'^2)}\nonumber\\
&+&C^{3}_{i}\langle\bar{d}d\rangle +C_{i}^5\langle\bar{d}\sigma_{\alpha \beta}
T^{a}G^{a\alpha\beta}d\rangle+\cdots,
\end{eqnarray}
where $\rho_i(s, s', q^2)$ is spectral density,
$C_i$ are the Wilson coefficients and $G^{a\alpha\beta}$
 is the gluon field strength tensor. We take for the strange quark condensate $\langle\overline{d}d\rangle=- (0.24\pm0.01)^3 ~GeV^3$ \cite{Prog12} and for the mixed quark-gluon condensate $\langle\bar{d}\sigma_{\alpha \beta}
T^{a}G^{a\alpha\beta}d\rangle=m_0^2\langle\overline{d}d\rangle$ with $m_0^2=(0.8\pm0.2)GeV^2$ \cite{Dosch}.

Furthermore, we make the usual assumption that the contributions of higher resonances are
well approximated by the perturbative expression
\begin{eqnarray}\label{eq214}
 -\frac{1}{4 \pi^2} \int_{s'_0}^{\infty}
ds'\int_{s_0}^{\infty} ds\frac{\rho_{i}(s, s',
q^2)}{(s-p^2)(s'-p'^2)},
\end{eqnarray}
with appropriate continuum thresholds $s_0$ and $s'_0$.

 The Cutkosky’s rule allows us to obtain the spectral densities of
 the correlation function for the Lorentz
structures appearing in the correlation function. The leading contribution
 comes from the perturbative term, shown
in Fig.\ref{F1}.
As a result, the spectral densities are obtained  to the double discontinuity in Eq.(\ref{eq214})
for vertices that are given in Appendix A.

\begin{figure}[th]
\begin{center}
\begin{picture}(50,10)
\put(-70,-40){ \epsfxsize=19cm \epsfbox{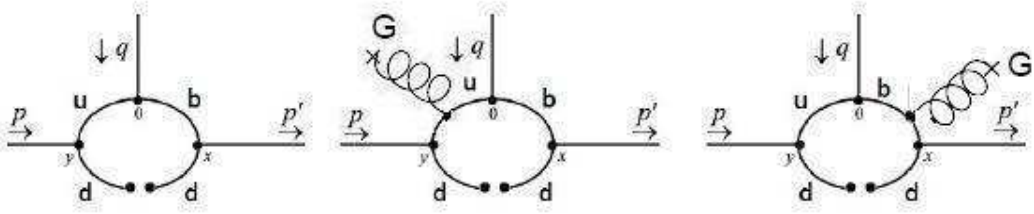}}
\end{picture}
\end{center}
\vspace*{1.85cm} \caption{Contribution of the
quark-quark and quark-gluon condensate for the bottom
off-shell.}\label{F2}
\end{figure}

We proceed to calculate the non-perturbative contributions in the QCD side that
contain the quark-quark and quark-gluon condensate. The quark-quark and quark-gluon condensate
is considered for when the light quark is
 a spectator \cite{Khodjamirian12},
Therefore only three important diagrams of dimension 3 and 5 remain from the non-perturbative part contributions when the bottom meson are off shell.
These diagrams named quark-quark and quark-gluon condensate are depicted in Fig.\ref{F2}.
For the pion off-shell,
there is no quark-quark and quark-gluon condensate contribution.

After some straightforward calculations  and applying the double Borel transformations with respect to
the $p^2(p^2\rightarrow M^2)$ and $p'^2(p'^2\rightarrow M'^2)$ as:
\begin{eqnarray}
{{B}}_{p^2}(M^2)(\frac{1}{p^2-m^2_u})^m=\frac{(-1)^m}{\Gamma(m)}
\frac{e^{-\frac{m_u^2}{M^2}}}{(M^2)^m}, \nonumber \\
{{B}}_{{p^{'}}^2}(M'^2)(\frac{1}{{p^{'}}^2-m^2_b})^n=\frac{(-1)^n}{\Gamma(n)}
\frac{e^{-\frac{m_b^2}{M'^2}}}{(M'^2)^n},
\end{eqnarray}
where $M^2$ and $M'^2$ are the Borel parameters,
the contribution of the quark-quark and quark-gluon condensate 
for the bottom meson off-shell case, are given by:
\begin{eqnarray}\label{eq218}
\Pi_{(non-per)}^{bottom}&=&\langle\overline{d}d\rangle~\frac{C^{bottom}}{M^4M'^4},
\end{eqnarray}
The explicit expressions for $C_{B_1B^*\pi[B_1B_0\pi(B_1B_1\pi)]}^{bottom}$ associated
with the $ B_1B^*\pi$, $ B_1B_0\pi$ and $ B_1B_1\pi$
vertices are given in Appendix B.

The gluon-gluon condensate is considered when the heavy quark is
 a spectator \cite{Likhoded}, and the bottom mesons are off-shell,
and there is no gluon-gluon condensate contribution.
Our numerical analysis shows that the contribution of the non-perturbative part containing
the quark-quark and quark-gluon diagrams is about $13\%$  and the gluon-gluon contribution
is about $4\%$  of the total and the main contribution comes from the perturbative
part of the strong form factors and we can ignore gluon-gluon contribution in our calculation.

The QCD sum rules for the strong form factors are obtained
after performing the Borel transformation with respect
to the variables $p^2 (B_{p^2}(M^2))$ and $p'^2(B_{p'}^2(M'^2))$
on the physical (phenomenological) and QCD parts and equating  these
two representations of the correlations, we obtain the corresponding equations for
the strong form factors as follows.

$\bullet$ For the $g_{B_1B^*\pi}(Q^2)$ form factors:
\begin{eqnarray}\label{eq219}
g^{B^*}_{B_1B^*\pi}(Q^2)&=&\frac{2(Q^2+m^2_{B^*})(m_u+m_d)}{m_{\pi}^2 m_{B^*}
f_{\pi}f_{B^*}f_{B_1}(m_{B_1}^2-m_{\pi}^2+Q^2)} ~e^{\frac{m_{\pi}^2}{M^2}}
e^{\frac{m_{B_1}^2}{M'^2}}
 \left\{-\frac{1}{4\pi^2}\int^{s'_0}_{(m_b
+m_d)^2} ds'\right. \nonumber \\ &&
\left. \times \int^{s_0}_{s_{1}} ds \rho^{B^*}(s,s',Q^2)
e^{-\frac{s}{M^2}} e^{-\frac{s'}{M'^2}}+ \langle d\bar d \rangle\frac{C_{B_1B^*\pi}^{B^*}}{M^2M'^2}
\right\},
\end{eqnarray}
\begin{eqnarray}\label{eq222}
g^{\pi}_{B_1B^*\pi}(Q^2)&=& \frac{2(Q^2+m^2_{\pi})(m_u+m_d)}{m_{\pi}^2 m_{B^*}
f_{\pi}f_{B^*}f_{B_1}(m_{B_1}^2-m_{B^*}^2+Q^2)} ~e^{\frac{m_{B^*}^2}{M^2}}
e^{\frac{m_{B_1}^2}{M'^2}}
 \left\{-\frac{1}{4\pi^2}\int^{s'_0}_{(m_b
+m_d)^2} ds'\right. \nonumber \\ &&
\left. \times \int^{s_0}_{s_{2}} ds \rho^{\pi}(s,s',Q^2)
e^{-\frac{s}{M^2}} e^{-\frac{s'}{M'^2}}\right\},
\end{eqnarray}

$\bullet$ For the $g_{B_1B_0\pi}(Q^2)$ form factors:
\begin{eqnarray}\label{eq220}
g^{B_0}_{B_1B_0\pi}(Q^2)&=& \frac{2(Q^2+m^2_{B_0})(m_u+m_d)}{m_{\pi}^2 m_{B_0}m_{B_1}
f_{\pi}f_{B_0}f_{B_1}(m_{B_1}^2+m_{\pi}^2+Q^2)} ~e^{\frac{m_{\pi}^2}{M^2}}
e^{\frac{m_{B_1}^2}{M'^2}}
 \left\{-\frac{1}{4\pi^2}\int^{s'_0}_{(m_b
+m_d)^2} ds'\right. \nonumber \\ &&
\left. \times \int^{s_0}_{s_{1}} ds \rho^{B_0}(s,s',Q^2)
e^{-\frac{s}{M^2}} e^{-\frac{s'}{M'^2}}+\langle d\bar d \rangle\frac{C_{B_1B_0\pi}^{B_0}}{M^2M'^2}
\right\},
\end{eqnarray}
\begin{eqnarray}\label{eq223}
g^{\pi}_{B_1B_0\pi}(Q^2)&=& \frac{2(Q^2+m^2_{\pi})(m_u+m_d)}{m_{\pi}^2 m_{B_0}m_{B_1}
f_{\pi}f_{B_0}f_{B_1}(m_{B_1}^2+m_{B_0}^2+Q^2)} ~e^{\frac{m_{B_0}^2}{M^2}}
e^{\frac{m_{B_1}^2}{M'^2}}
 \left\{-\frac{1}{4\pi^2}\int^{s'_0}_{(m_b
+m_d)^2} ds'\right. \nonumber \\ &&
\left. \times \int^{s_0}_{s_{2}} ds \rho^{\pi}(s,s',Q^2)
e^{-\frac{s}{M^2}} e^{-\frac{s'}{M'^2}}\right\},
\end{eqnarray}

$\bullet$ For the $g_{B_1B_1\pi}(Q^2)$ form factors:
\begin{eqnarray}\label{eq221}
g^{B_1}_{B_1B_1\pi}(Q^2)&=& -i \frac{(Q^2+m^2_{B_1})(m_u+m_d)}{m_{\pi}^2 m_{B_1}^2
f_{\pi}f_{B_1}^2} ~e^{\frac{m_{\pi}^2}{M^2}}
e^{\frac{m_{B_1}^2}{M'^2}}
 \left\{-\frac{1}{4\pi^2}\int^{s'_0}_{(m_b
+m_d)^2} ds'\right. \nonumber \\ &&
\left. \times \int^{s_0}_{s_{1}} ds \rho^{B_1}(s,s',Q^2)
e^{-\frac{s}{M^2}} e^{-\frac{s'}{M'^2}}+\langle d\bar d \rangle\frac{C_{B_1B_1\pi}^{B_1}}{M^2M'^2}
\right\},
\end{eqnarray}

\begin{eqnarray}\label{eq224}
g^{\pi}_{B_1B_1\pi}(Q^2)&=& -i\frac{(Q^2+m^2_{\pi})(m_u+m_d)}{m_{\pi}^2 m_{B_1}^2
f_{\pi}f_{B_1}^2} ~e^{\frac{m_{B_1}^2}{M^2}}
e^{\frac{m_{B_1}^2}{M'^2}}
 \left\{-\frac{1}{4\pi^2}\int^{s'_0}_{(m_b
+m_d)^2} ds'\right. \nonumber \\ &&
\left. \times \int^{s_0}_{s_{2}} ds \rho^{\pi}(s,s',Q^2)
e^{-\frac{s}{M^2}} e^{-\frac{s'}{M'^2}}\right\},
\end{eqnarray}
where $Q^2=-q^2$, $ s_0 $ and $s'_0  $ are the continuum
thresholds and $s_1$ and $s_2$ are the lower limits of the integrals over $s$ as:
\begin{eqnarray}\label{eq225}
s_{1(2)}=\frac{(m_{d(b)}^{2}+q^2-m_{u}^{2}-s')
(m_{u}^{2}s'-q^2m_{d(b)}^{2})}{(m_{u}^{2}-q^2)(m_{d(b)}^{2}-s')}~.
\end{eqnarray}
\section{NUMERICAL ANALYSIS}
In this section, the expressions of QCD sum
rules obtained for the considered strong coupling constants are investigated.
We choose the values of the meson and quark masses as:
$m_u=(1.7-3.3)~MeV$, $m_d = (3.5-6.0)~MeV$, 
$m_{\pi}=14~MeV$, $m_{B^*}=5.32~GeV$, $m_{D^*}=2.01~GeV$, 
$m_{B_1}=5.72~GeV$, $m_{D_1}=2.42~GeV$, 
$m_{B_0}=5.70~GeV$, $m_{D_0}=2.36~GeV$.
Also the leptonic decay constants used in this calculation are taken as:
$f_{\pi}=130.41~MeV$\cite{pdg1}, 
$f_{B^*}=238\pm10~MeV$, $f_{D^*}=340\pm12~MeV$\cite{Wang2},
$f_{B_1}= 196.9 \pm 8.9~ MeV$, $f_{D_1}= 218.9 \pm 11.3~MeV$\cite{Bazavov},
$f_{B_0}=280\pm31~MeV$, $f_{D_0}=334 \pm 8.6~MeV$\cite{Huang2}.
For a comprehensive analysis of the strong coupling constants,
we use the following values of the quark masses $m_b$ and $m_c$
in two sets: setI, $m_b(\overline{MS})=4.67~GeV$\cite{Beringer1}, $m_c=1.26~GeV$\cite{Janbazi, Richard} and set II, $m_b(1S)=4.19~GeV$\cite{Beringer1}, $m_c=1.47~GeV$\cite{Janbazi, Richard}.

The expressions for the strong form factors in  Eqs.(\ref{eq219}-\ref{eq224}) should not
 depend on the Borel variables $M^2$
and $M'^2$.
Therefore, one has to work in a region
where the approximations made are supposedly acceptable and where the result depends only
moderately on the Borel variables.
In this work we use the following relations between the Borel masses
$M ^2$ and $ M'^2 $ \cite{MChiapparini, Rodrigues3}: $ \frac{M^2}{M'^2}=\frac{m_{\pi}^2}{m_{B_1}^2-m_{b}^2} $
for bottom meson off-shell
and $M^2=M'^2 $
for pion meson  off-shell. The
values of the continuum thresholds $s_0=(m+\Delta)^2$ and
$s'_0=(m_{B_1}+\Delta)^2$, where m is the $\pi$ mass, for $B^*[B_0(B_1)]$ off-shell and the
$B^*[B_0(B_1)]$ meson mass, for $\pi$ off-shell and $ \Delta $ varies between: $ 0.4~GeV \leq\Delta\leq 1~GeV$ \cite{Janbazi, Richard}.

\begin{figure}[th]
\begin{center}
\begin{picture}(110,26)
\put(-20,-20){ \epsfxsize=7.5cm \epsfbox{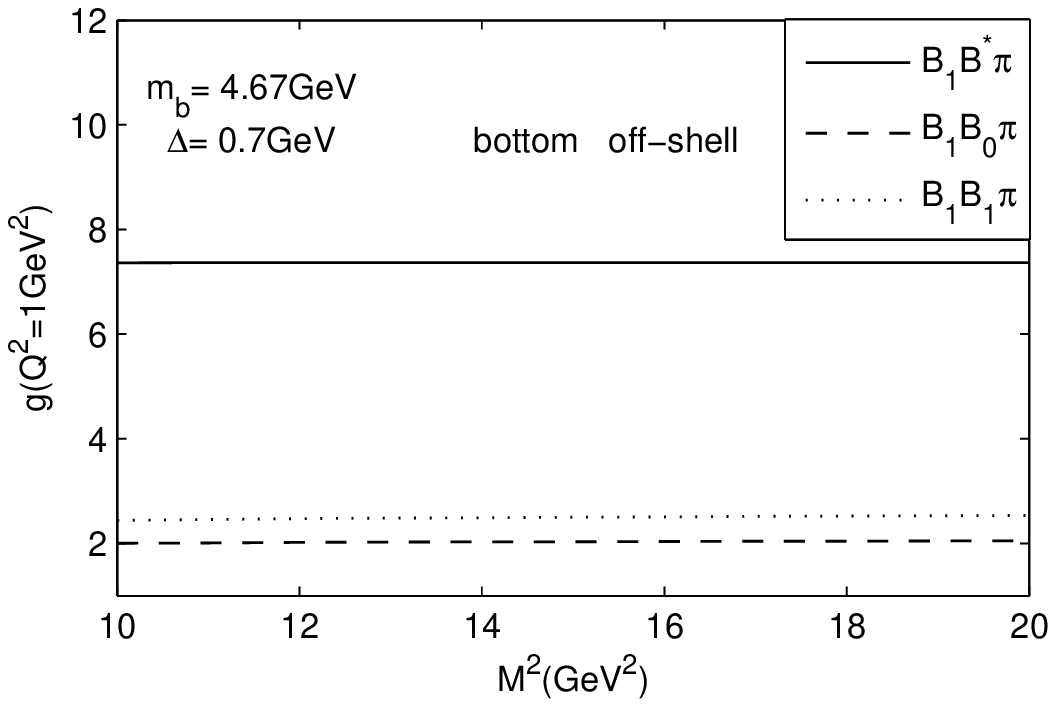} \epsfxsize=7.5cm \epsfbox{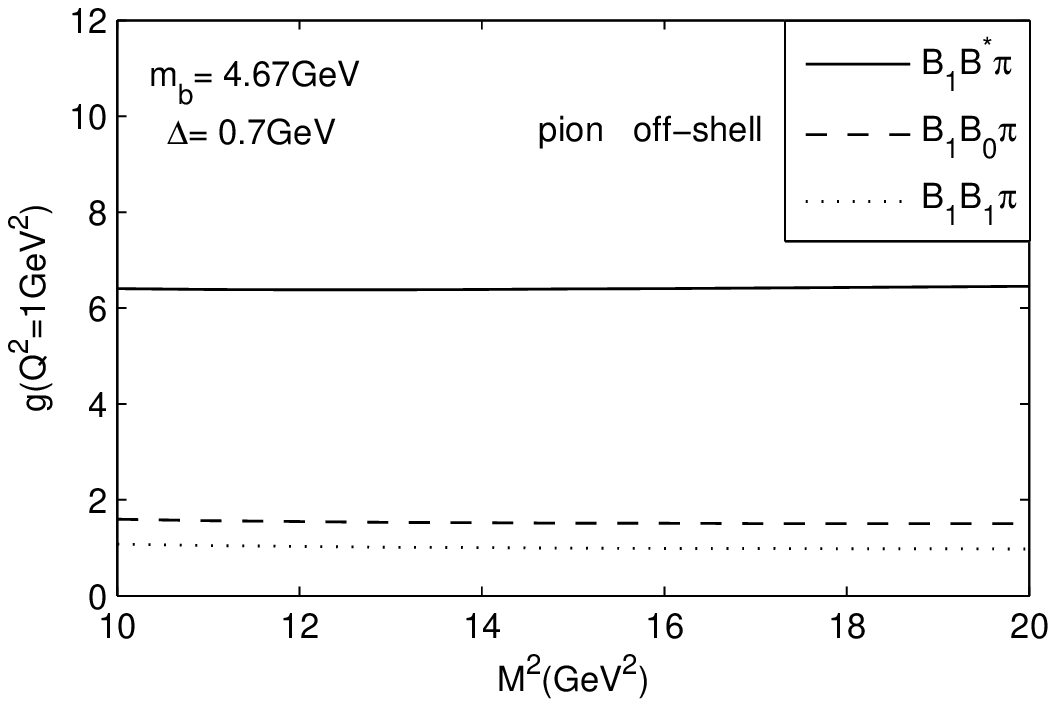} }
\end{picture}
\end{center}
\vspace*{1.5cm} \caption{The strong form factors $g_{B_1B^*\pi}$,  $g_{B_1B_0\pi}$ and $g_{B_1B_1\pi}$ as functions of the Borel mass parameter $M^2$ with
for two cases bottom off-shell
meson (left) and pion off-shell
mesons (right).}\label{F31}
\end{figure}

\begin{figure}
\vspace{0cm}
\includegraphics[width=5.1cm,height=5cm]{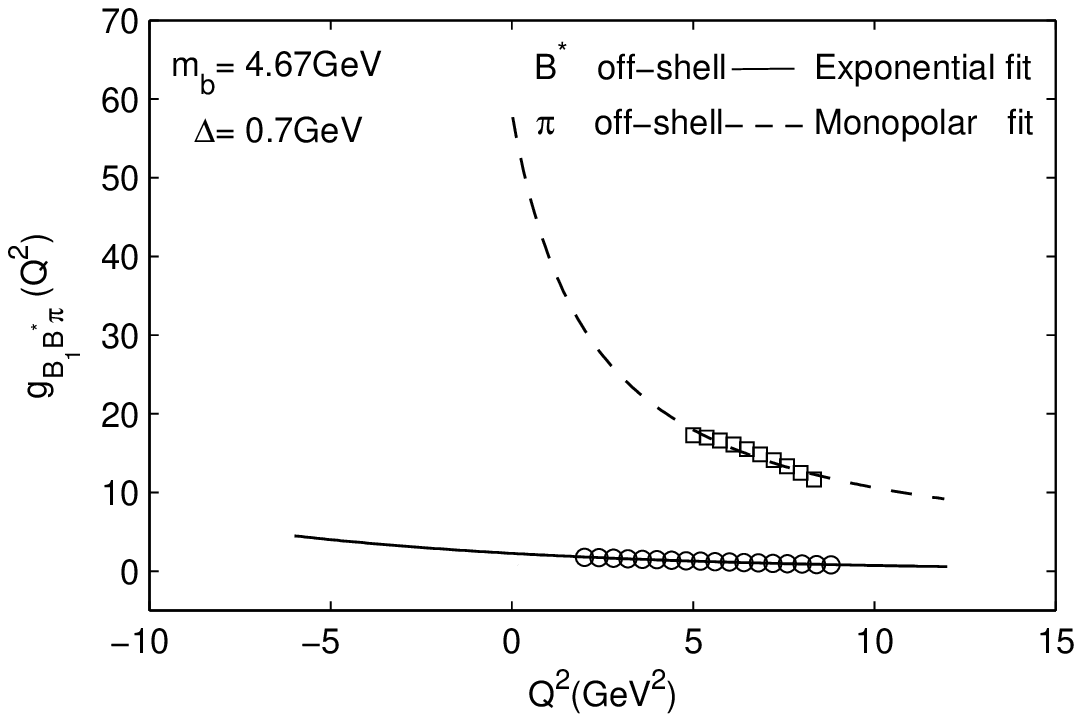}
\hspace{0.5cm}
\includegraphics[width=5.1cm,height=5cm]{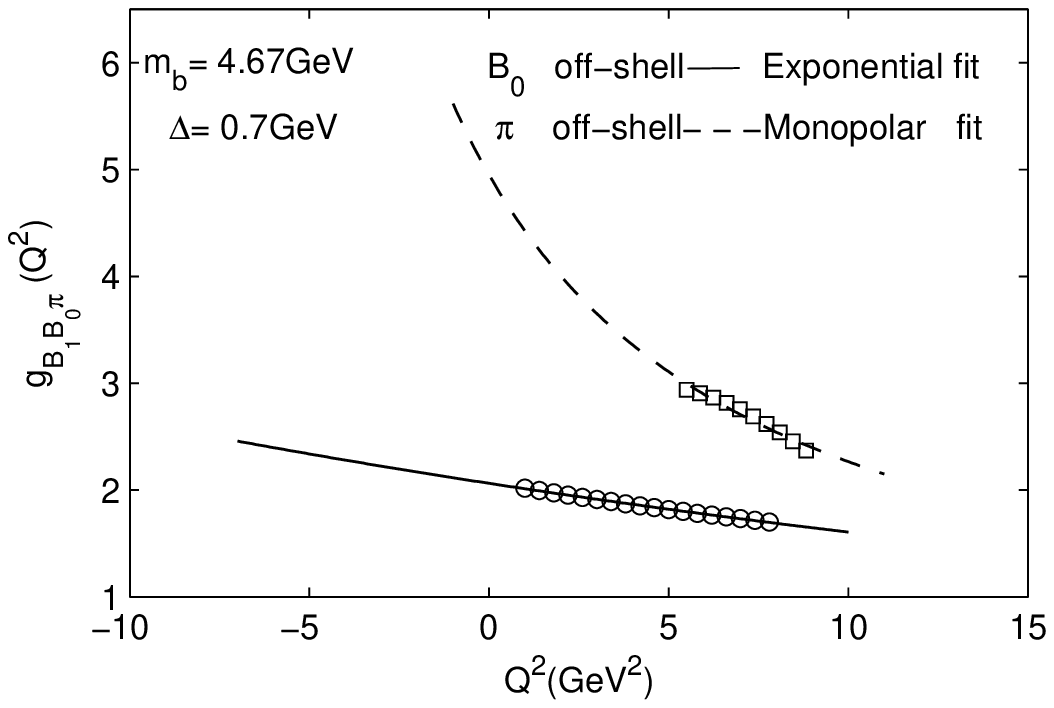}
\vspace{1.cm}
\includegraphics[width=5.1cm,height=5cm]{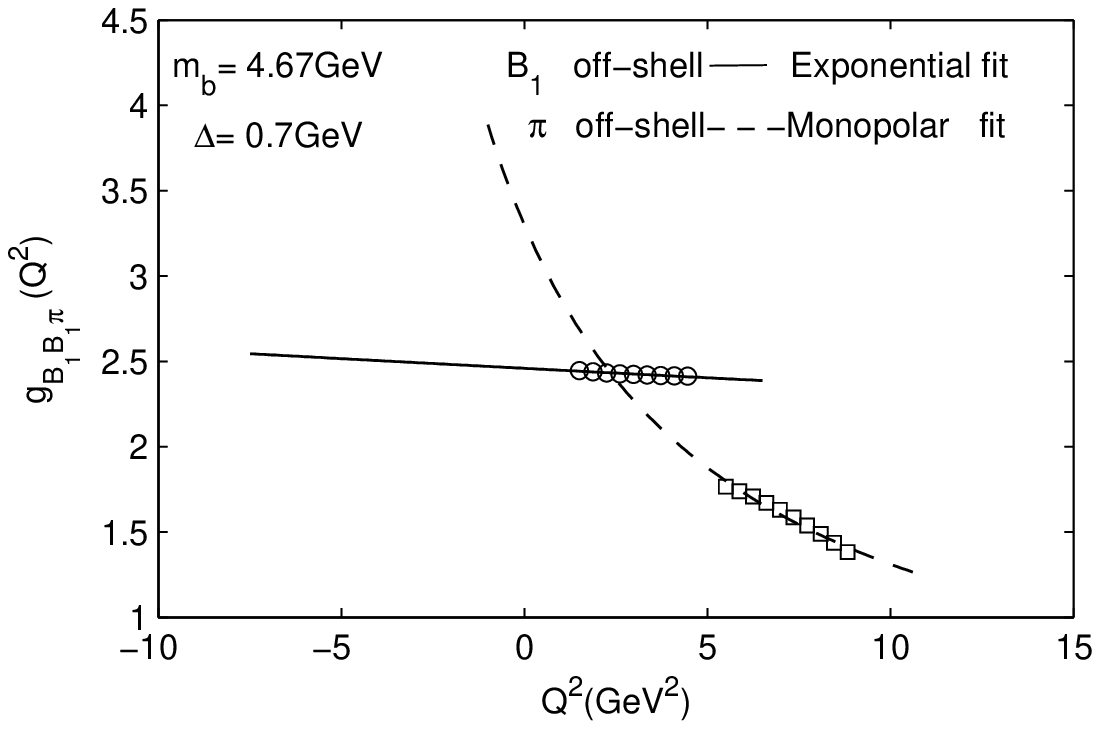}
\vspace{-7ex}
\caption{The strong form factors $g_{B_1B^*\pi}$,
  $g_{B_1B_0\pi}$ and $g_{B_1B_1\pi}$ on $Q^2$ for the bottom off-shell and  the pion off-shell mesons. The small circles and boxes correspond to the form factors via the 3PSR calculations.}
\label{F32}
\end{figure}

Using $\Delta=0.7GeV$, $m_b=4.67~GeV$
and fixing $Q^2 = 1GeV^2$, We found a good stability of the sum rule in the interval
 $10~GeV^2\leq M ^2\leq20~GeV^2$
for the two cases of bottom and pion being off-shell. The dependence of the strong form factors $g_{B_1B^*\pi}$,  $g_{B_1B_0\pi}$ and $g_{B_1B_1\pi}$ on Borel mass parameters for off-shell bottom and pion mesons are shown in Fig.\ref{F31}.
We have chosen the Borel mass to be $ M^2= 13~GeV^2 $.
Having determined $M^2 $, we calculated the $Q^2$
dependence of the form factors. We present the results in
Fig.\ref{F32} for the $g_{B_1B^*\pi}$,  $g_{B_1B_0\pi}$ and $g_{B_1B_1\pi}$ vertices.
In this figures, the small
circles and boxes correspond to the form factors in the interval where the
sum rule is valid. As it is seen, the form factors and their fit
functions coincide together, well.

We discuss a difficulty inherent to the calculation of coupling
constants with QCDSR. The solution of Eqs.(\ref{eq219}-\ref{eq224}) is numerical and restricted to a singularity-free region in
the $Q^2 $ axis, usually located in the space-like region. Therefore, in order to reach the pole position,
$Q^2 = -m_m^2$, we must fit the solution, by finding a function $g(Q^2)$ which is then extrapolated to
the pole, yielding the coupling constant.

The uncertainties associated with the extrapolation procedure, for each vertex is minimized by
performing the calculation twice, first  putting one meson and then another meson off-shell, to obtain two
form factors $g^{bottom}$ and   $g^{pion}$ and equating these two functions
at the respective poles. The superscripts in parenthesis indicate which meson is off-shell.
In order to reduce the freedom in the extrapolation and constrain the form factor, we calculate
and fit simultaneously the values of $g(Q^2)$  with the pion off-shell. We tried to fit our results to a monopole form, since this is often used for form factors \cite{99}.

For the off-shell pion meson, Our numerical
calculations show that the sufficient parametrization of the form
factors with respect to $Q^2$ is:
\begin{eqnarray}\label{eq31}
g(Q^2)=\frac{A}{Q^2+B},
\end{eqnarray}
and for off-shell bottom meson the  strong form factors can be
fitted by the exponential fit function as given:
\begin{eqnarray}\label{eq32}
g(Q^2)=A~e^{-Q^2/B}.
\end{eqnarray}

\begin{table}[th]
\caption{Parameters appearing in the fit functions for the $g_{B_1B^*\pi}$,
  $g_{B_1B_0\pi}$ and $g_{B_1B_1\pi}$ vertices for $\Delta_1=0.7 ~GeV$ and $m_b(\overline{MS})=4.67~GeV$ (set I) and $m_b(1S)=4.19~GeV$ (set II).}\label{T31}
\begin{ruledtabular}
\begin{tabular}{ccccc}
&$\mbox{set I}$&$$&$\mbox{set II}$&\\
\hline
$\mbox{Form factor}$&$A$&$B$&$A$&$B$\\
$g^{B^*}_{B_1B^*\pi}$& 2.26&8.73&4.35&11.56\\
$g^{\pi}_{B_1B^*\pi}$&129.87&2.23&301.25&6.12 \\
$g^{B_0}_{B_1B_0\pi}$&2.06&39.93&2.47&37.53\\
$g^{\pi}_{B_1B_0\pi}$&41.77&8.44&308.03&54.43  \\
$g^{B_1}_{B_1B^*\pi}$&2.46&219.04&2.59&132.90 \\
$g^{\pi}_{B_1B_1\pi}$&21.77&6.60&205.82&60.51
\end{tabular}
\end{ruledtabular}
\end{table}

\begin{table}[th]
\caption{The strong coupling constants $g_{B_1B^*\pi}$,  $g_{B_1B_0\pi}$ and
$g_{B_1B_1\pi}$ . }\label{T32}
\begin{ruledtabular}
\begin{tabular}{ccccccc}
&$\mbox{set I}$&$$&$\mbox{set II}$&$$\\
$\mbox{Coupling constant}$&$\mbox{bottom-off-sh}$&$\mbox{pion-off-sh}$&$\mbox{bottom-off-sh}$&$\mbox{pion-off-sh}$&$\mbox{Average}$ \\
\hline
$g_{B_1B^*\pi}$&$57.63\pm15.53$&$58.72\pm15.43$&$50.32\pm13.24$&$49.38\pm14.26$&$54.01\pm15.51$\\
$g_{B_1B_0\pi}$&$4.68\pm1.44$&$4.96\pm1.08$&$5.87\pm1.34$&$5.66\pm1.13$&$5.29\pm1.40$\\
$g_{B_1B_1\pi}(GeV^{-1})$&$2.86\pm0.43$&$3.31\pm0.27$&$3.31\pm0.25$&$3.89\pm0.18$&$3.57\pm0.53$
\end{tabular}
\end{ruledtabular}
\end{table}

The values of the parameters $A$ and $B$ are given in the Table
\ref{T31}. We define the coupling constant as the value of the strong
coupling form factor at $Q^2 = -m_m^2$ in the Eq. (\ref{eq31}) and
Eq. (\ref{eq32}), where $m_m$ is the mass of the off-shell meson. Considering the uncertainties result with the continuum threshold and uncertainties
in the values of the other input parameters, we obtain the average values of the strong coupling
constants in different sets shown in Table
\ref{T32}.

We can see that for the two cases considered here, the off-
shell  bottom and pion meson,  give compatible results for the
coupling constant.

The same method described in section II with little change in the
containing perturbative and non-perturbative parts,  where $\rho_{D_1D^*\pi[D_1D_0\pi(D_1D_1\pi)]}^{charm(pion)}=\rho_{B_1B^*\pi[B_1B_0\pi(B_1B_1\pi)]}^{bottom(pion)}{|}_{b\rightarrow c}$, $C^{charm}_{D_1D^*\pi[D_1D_0\pi(D_1D_1\pi)]}=C_{B_1B^*\pi[B_1B_0\pi(B_1B_1\pi)]}^{bottom}{|}_{b\rightarrow c}$, we can easily find similar results in  Eqs.(\ref{eq219}-\ref{eq224})
for strong form factors $g_{D_1D^*\pi}$,  $g_{D_1D_0\pi} $ and $g_{D_1D_1\pi}$ and also use the following relations between the Borel masses
$M ^2$ and $ M'^2 $ : $ \frac{M^2}{M'^2}=\frac{m_{\pi}^2}{m_{D_1}^2-m_{c}^2} $
for charm meson off-shell
and $M^2=M'^2 $
for pion meson  off-shell. The
values of the continuum thresholds $s_0=(m+\Delta)^2$ and
$s'_0=(m_{D_1}+\Delta)^2$, where m is the $\pi$ mass, for $D^*[D_0(D_1)]$ off-shell and the
$D^*[D_0(D_1)]$ meson mass, for the $\pi$ off-shell and $ \Delta $ being between $ 0.4~GeV \leq\Delta\leq 1~GeV$.

\begin{figure}[th]
\begin{center}
\begin{picture}(110,20)
\put(-20,-25){ \epsfxsize=7.5cm \epsfbox{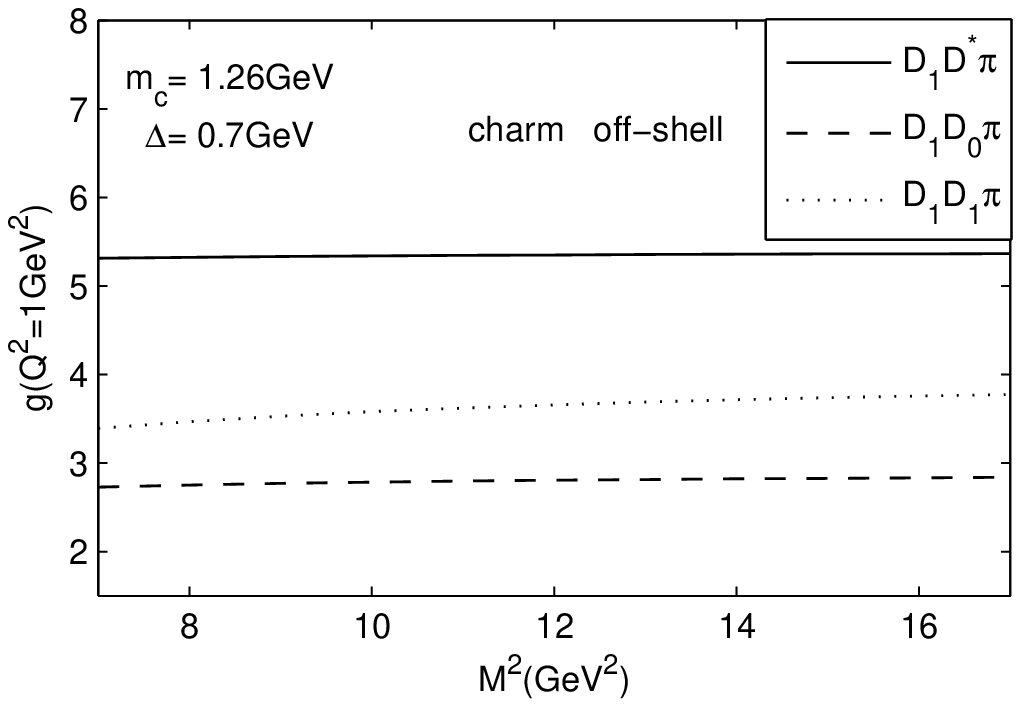} \epsfxsize=7.5cm \epsfbox{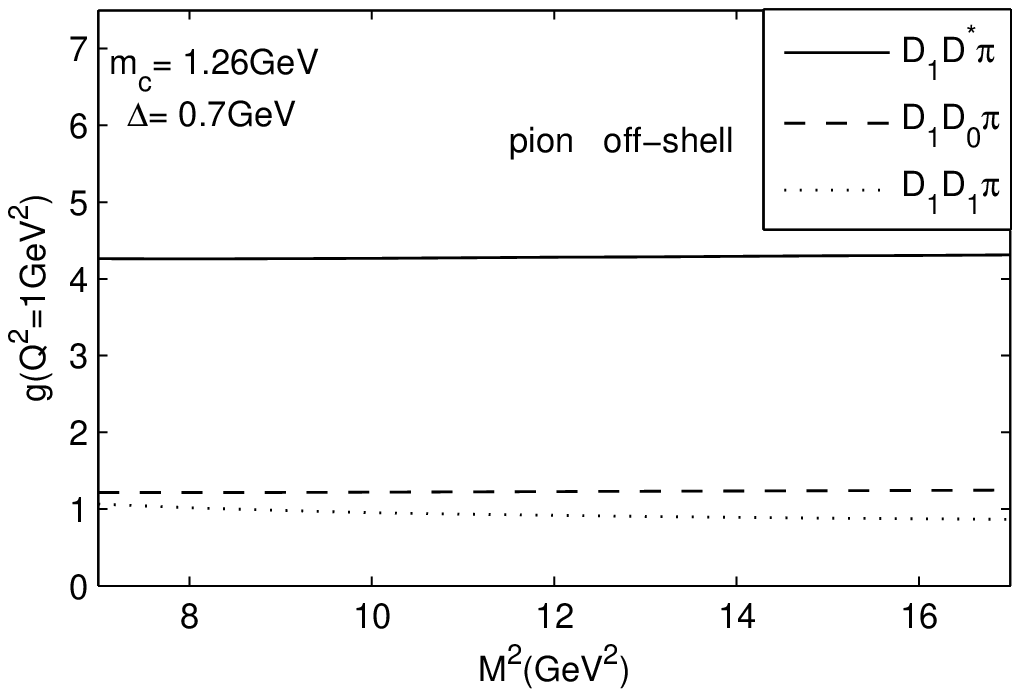} }
\end{picture}
\end{center}
\vspace*{1.5cm} \caption{The strong form factors $g_{D_1D^*\pi}$,  $g_{D_1D_0\pi} $ and $g_{D_1D_1\pi}$ as functions of the Borel mass parameter $M^2$ 
for the two cases of charm off-shell
meson (left), and pion off-shell
meson (right).}\label{F33}
\end{figure}

\begin{figure}
\vspace{0cm}
\includegraphics[width=5.1cm,height=5cm]{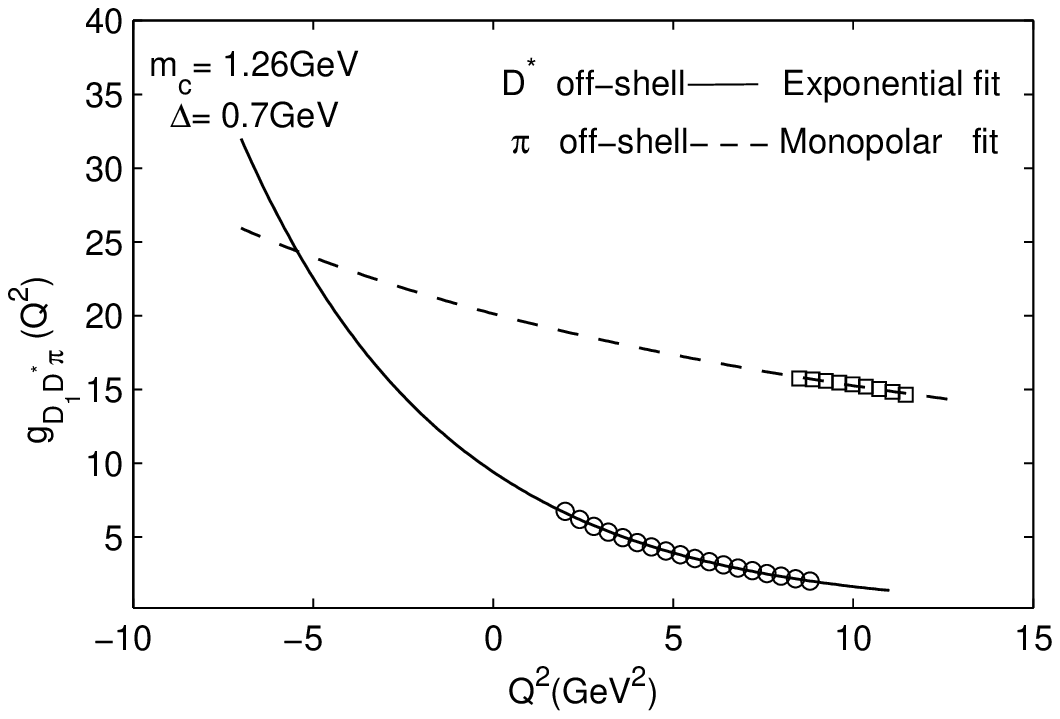}
\hspace{0.5cm}
\includegraphics[width=5.1cm,height=5cm]{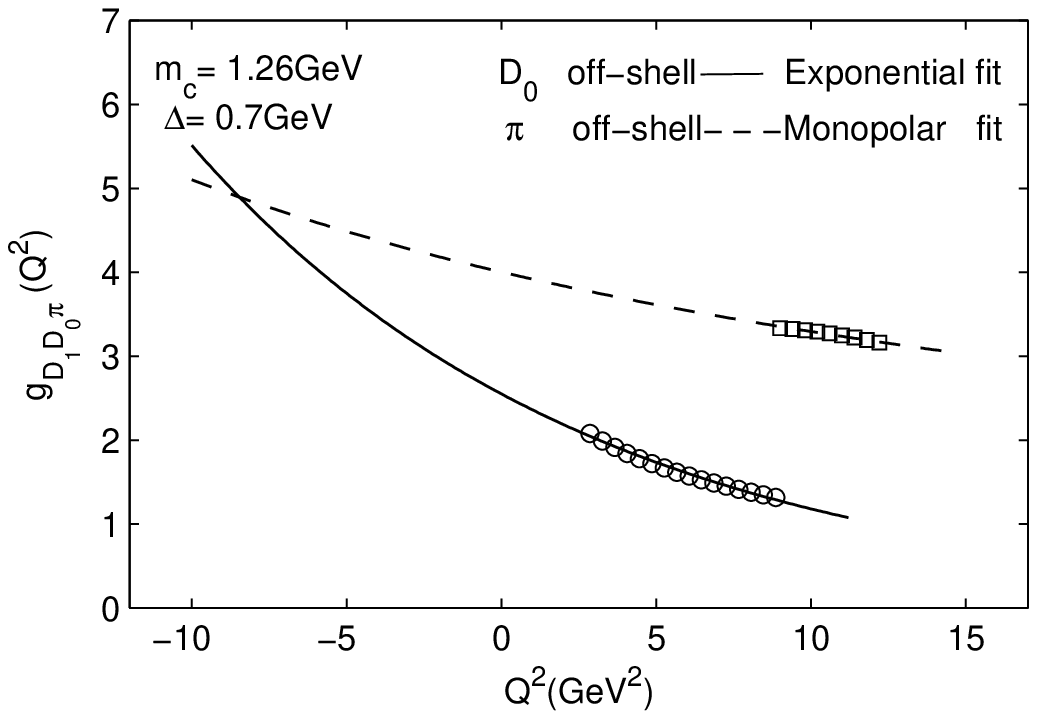}
\vspace{1.cm}
\includegraphics[width=5.1cm,height=5cm]{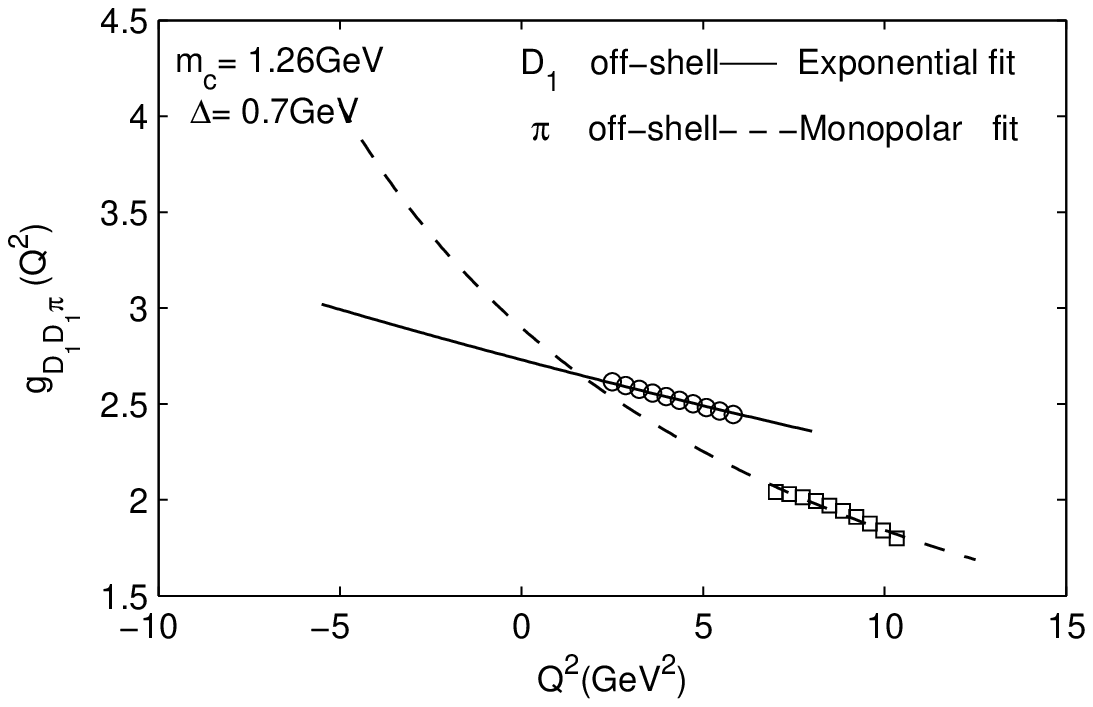}
\vspace{-7ex}
\caption{The strong form factors $g_{D_1D^*\pi}$,  $g_{D_1D_0\pi} $ and $g_{D_1D_1\pi}$ dependence
on $Q^2$ for the charm off-shell and  the pion off-shell mesons. The small circles and boxes correspond to the form factors via the 3PSR calculations.}
\label{F34}
\end{figure}

Using $\Delta=0.7GeV$, $m_c=1.26~GeV$
and fixing $Q^2 = 1GeV^2$, we found a good stability of the sum rule in the interval
 $7~GeV^2\leq M^2\leq17~GeV^2$
for two cases of charm and pion off-shell. The dependence of the strong form factors $g_{D_1D^*\pi}$,  $g_{D_1D_0\pi}$ and $g_{D_1D_1\pi}$ on Borel mass parameters for the off-shell charm and pion mesons are shown in Fig.\ref{F33}.
We have chosen the Borel mass to be $ M^2= 10~GeV^2 $.
Having determined $M^2 $, we calculated the $Q^2$
dependence of the form factors. We present the results in
Fig.\ref{F34} for the $g_{D_1D^*\pi}$,  $g_{D_1D_0\pi} $ and $g_{D_1D_1\pi}$ vertices.

The dependence of the above strong form factors on $Q^2$ to the
full physical region is estimated, using Eq.(\ref{eq31}) and Eq.(\ref{eq32}) for the pion and charm
off-shell mesons, respectively. 
The values of the parameters $A$ and $B$ are given in the Table
\ref{T33}.

Considering the uncertainties result with the continuum threshold and uncertainties
in the values of the other input parameters, we obtain the average values of the strong coupling
constants in different values of the different sets shown in Table
\ref{T34}. 

\begin{table}[th]
\caption{Parameters appearing in the fit functions for the $g_{D_1D^*\pi}$,
  $g_{D_1D_0\pi}$ and $g_{D_1D_1\pi}$ vertices for $\Delta_1=0.7 ~GeV$ and $m_c=1.26$ (set I) and $m_c=1.47$ (set II).}\label{T33}
\begin{ruledtabular}
\begin{tabular}{ccccc}
&$\mbox{set I}$&$$&$\mbox{set II}$&\\
\hline
$\mbox{Form factor}$&$A$&$B$&$A$&$B$\\
$g^{D^*}_{D_1D^*\pi}$& 9.41&5.72&9.58&5.83\\
$g^{\pi}_{D_1D^*\pi}$&63.07&31.30&86.40&4.18 \\
$g^{D_0}_{D_1D_0\pi}$&2.55&12.97&2.37&13.05\\
$g^{\pi}_{D_1D_0\pi}$&185.69&46.40&32.98&8.49  \\
$g^{D_1}_{D_1D^*\pi}$&2.75&49.54&2.21&14.40 \\
$g^{\pi}_{D_1D_1\pi}$&50.54&17.44&13.79&3.92
\end{tabular}
\end{ruledtabular}
\end{table}
\begin{table}[th]
\caption{The strong coupling constants $g_{D_1D^*\pi}$,  $g_{D_1D_0\pi}$ and
$g_{D_1D_1\pi}$ . }\label{T34}
\begin{ruledtabular}
\begin{tabular}{cccccc}
&$\mbox{set I}$&$$&$\mbox{set II}$&$$\\
$\mbox{Coupling constant}$&$\mbox{charm-off-sh}$&$\mbox{pion-off-sh}$&$\mbox{charm-off-sh}$&$\mbox{pion-off-sh}$&$\mbox{Average}$\\
\hline
$g_{D_1D^*\pi}$&$19.07\pm4.21$&$20.14\pm4.49$&$19.16\pm3.87$&$20.77\pm3.92$&$19.78\pm3.32$\\
$g_{D_1D_0\pi}$&$3.92\pm0.93$&$4.03\pm1.01$&$3.63\pm0.84$&$3.89\pm0.73$&$3.87\pm0.86$\\
$g_{D_1D_1\pi}(GeV^{-1})$&$3.09\pm0.63$&$2.90\pm0.52$&$3.31\pm0.54$&$3.54\pm0.61$&$3.21\pm0.49$
\end{tabular}
\end{ruledtabular}
\end{table}

In Table \ref{T35} we compare our obtained values, with the findings of others, previously calculated. From this Table we see that our result of the coupling constants is in a fair agreement with the calculations in
refs.\cite{Colangelo,Hungchong,Zhu}.

\begin{table}[th]
\caption{Comparison of our results with the other published results. The results of Refs.  \cite{Colangelo,Aliev} are from light-cone QCD sum rules, the result from Ref.\cite{Hungchong} is from the QCD sum rules and the short distance
expansion, and the result of Ref.\cite{Zhu} is from the light-cone QCD sum rules
in HQET.}
\label{T35}
\begin{ruledtabular}
\begin{tabular}{ccccccc}
 & $g_{B_1B^*\pi}$ &$g_{B_1B_0\pi}$&$g_{B_1B_1\pi}(GeV^{-1})$&$g_{D_1D^*\pi}$&$g_{D_1D_0\pi}$&$g_{D_1D_1\pi}(GeV^{-1})$
\\ \hline
Our result &$54.01\pm15.51$& $5.29\pm1.40$& $3.57\pm0.53$& $19.78\pm 3.32$& $3.87\pm 0.86$&$3.21\pm0.49$\\
Ref.\cite{Colangelo}&$56\pm15$& $5.39\pm2.15$&-&$23\pm 5$& $3.43\pm1.37$& -\\
Ref.\cite{Hungchong}&-&-&-&$19.12\pm2.42$&-& $2.59\pm 0.61$\\
Ref.\cite{Aliev}&$68.64\pm8.58$& -&-& $12.10\pm2.42$&-&-\\
Ref.\cite{Zhu} &$58.89\pm9.81$& $4.73\pm1.14$&$2.60\pm0.60$& -& -& -
\end{tabular}
\end{ruledtabular}
\end{table}

\section{conclusion}
In this article, we analyzed the vertices $B_1B^*\pi$,
 $B_1B_0\pi$, $B_1B_1\pi$, $D_1D^*\pi$, $D_1D_0\pi$
 and $D_1D_1\pi$ within
the framework of the three point QCD sum rules 
approach in an unified way. The
strong coupling constants could give useful information 
about strong interactions of the  strange bottomed and strange
charmed mesons and also are important ingredients for 
estimating the absorption cross section of the
$J/\psi$ by the $\pi$ mesons.

\appendix
\begin{center}
{ \textbf{Appendix A: PERTURBATIVE CONTRIBUTIONS }}
\end{center}
\setcounter{equation}{0} \renewcommand{\theequation}

In this appendix, The perturbative contributions for the sum rules defined in Eqs.(\ref{eq219}-\ref{eq224}) are:
\begin{eqnarray*}
\rho^{B^*(\pi)}_{B_1B^*\pi} &=& 4 N_c I_0 k\left[2A\left(m_1-m_{3(2)}\right)-m_1m_2m_3+m_2m_3^2+m_3^3-m_1m_3^2-\frac{\Delta}{2}\left(m_2+m_3\right)\right. \nonumber \\ &&
\left.+\frac{\Delta'}{2}\left(m_1-m_3\right)+\frac{m_3u}{2}\right]
,\nonumber \\
\rho^{B_0(\pi)}_{B_1B_0\pi}&=& 4 N_c I_0\left[B_2\left(m_2m_3-km_1m_2+km_1m_3-m_3^2+\Delta-\frac{u}{2}\right)+km_3^2-m_3m_1-k\frac{\Delta}{2}\right],
\nonumber \\
\rho^{B_1(\pi)}_{B_1B_1\pi}&=& 4i N_c I_0\left[B_1\left(m_3-km_1\right)+B_2\left(m_2+m_3\right)+m_3\right],
\end{eqnarray*}
The explicit expressions of the coefficients in the spectral densities
entering the sum rules are given as:
\begin{eqnarray*}
I_0(s,s',q^2) &=& \frac{1}{4\lambda^\frac{1}{2}(s,s',q^2)},\nonumber \\
\Delta &=& (s+m_3^2-m_1^2),\nonumber \\
\Delta' &=& (s'+m_3^2-m_2^2),\nonumber \\
u &=& s+s'-q^2,\nonumber \\
\lambda(s,s',q^2) &=& s^2+ s'^2+ q^4- 2sq^2- 2s'q^2- 2ss',\nonumber \\
A &=& -\frac{1}{2\lambda(s,s',q^2)}[4ss'm_3^2-s\Delta'^2-s'\Delta^2-u^2m_3^2+u\Delta\Delta'],\nonumber \\
B_1 &=& \frac{1}{\lambda(s,s',q^2)} \left [2 s' \Delta -\Delta' u\right],\nonumber \\
B_2 &=& \frac{1}{\lambda(s,s',q^2)} \left [2 s \Delta' -\Delta u\right],
\end{eqnarray*}
Where $k=1$, $m_1=m_u$, $m_2=m_b$, $m_3=m_d$ for bottom meson off-shell and $k=-1$, $m_1=m_u$, $m_2=m_d$, $m_3=m_b$ for pion meson off-shell,
$N_c=3$ represents the color factor.
\vspace*{.7cm}
\appendix
\begin{center}
{ \textbf{Appendix B: NON-PERTURBATIVE CONTRIBUTIONS }}
\end{center}
\setcounter{equation}{0} \renewcommand{\theequation}

In this appendix,  the explicit expressions of the coefficients of
the quark-quark and quark-gluon  condensate of the strong form
factors for the vertices $B_1B^*\pi$, $B_1B_0\pi$ and $B_1B_1\pi$
with applying the double Borel transformations
are given.
\begin{eqnarray*}
C_{B_1B^*\pi}^{B^*} &=&(\frac{7M^2m_b^2m_0^2}{24}-\frac{M^2M'^2m_0^2}{6}+\frac{M'^2m_b^2m_0^2}{8}-\frac{m_0^2m_b^4}{8}-\frac{M^2M'^2m_bm_d}{4}+\frac{M^2M'^2m_b^2}{2}\\&&-\frac{M^2m_b^3m_d}{4}+\frac{M^2m_bm_dq^2}{4}-\frac{M^2m_b^2m_d^2}{2}-\frac{M^2m_0^2m_bm_u}{4}-\frac{3M'^2m_0^2m_bm_u}{4}\\&&-M^2M'^2m_bm_u+\frac{m_0^2m_b^3m_u}{4}+\frac{M'^2m_0^2m_b^3m_u}{2M^2}-\frac{M^2M'^2m_dm_u}{4}+\frac{M^2m_b^2m_dm_u}{2}\\&&+\frac{M'^2m_b^2m_dm_u}{4}+\frac{M^2m_bm_d^2m_u}{2}+\frac{M'^2m_bm_d^2m_u}{2}-\frac{m_b^3m_d^2m_u}{2}+\frac{M^2m_0^2m_u^2}{24}\\&&+\frac{M'^2m_0^2m_u^2}{4}+\frac{M^2M'^2m_u^2}{2}-\frac{m_0^2m_b^2m_u^2}{4}-\frac{M'^2m_bm_dm_u^2}{2}-\frac{M'^2m_d^2m_u^2}{2}+\frac{m_b^2m_d^2m_u^2}{2}\\&&+\frac{m_0^2m_bm_u^3}{4}-\frac{M'^2m_0^2m_bm_u^3}{2M^2}+\frac{M'^2m_dm_u^3}{4}-\frac{m_bm_d^2m_u^3}{2}-\frac{M'^2m_dm_uq^2}{4}-\frac{7M^2m_0^2q^2}{24}\\&&-\frac{3M'^2m_0^2q^2}{8}-\frac{M^2M'^2q^2}{2}+\frac{m_0^2m_b^2q^2}{4}+\frac{M^2m_d^2q^2}{2}+\frac{M'^2m_d^2q^2}{2}-\frac{m_b^2m_d^2q^2}{2}\\&&-\frac{m_0^2m_bm_uq^2}{4})\times
e^{-\frac{m_u^2}{M^2}}~e^{-\frac{m_b^2}{M'^2}},
\end{eqnarray*}
\begin{eqnarray*}
C_{B_1B_0\pi}^{B_0} &=&(\frac{M^2m_0^2m_b}{4}-\frac{M^2m_b^2m_d}{2}-\frac{M^2m_bm_d^2}{2}-\frac{3m_0^2M'^2m_u}{4}-M^2M'^2m_u+\frac{m_0^2m_b^2m_u}{4}\\&&-\frac{M^2m_bm_dm_u}{2}+\frac{M^2m_d^2m_u}{2}-\frac{m_b^2m_d^2m_u}{2}-\frac{M^2m_dm_u^2}{2}-\frac{M'^2m_dm_u^2}{2}+\frac{M'^2m_d^2mu}{2}\\&&+\frac{m_0^2m_u^3}{4}-\frac{m_d^2m_u^3}{2}+\frac{M^2m_dq^2}{2}-\frac{m_0^2m_uq^2}{4}+\frac{m_d^2m_uq^2}{2})\times
e^{-\frac{m_u^2}{M^2}}~e^{-\frac{m_b^2}{M'^2}},
\end{eqnarray*}
\begin{eqnarray*}
C_{B_1B_1\pi}^{B_1} &=&i(\frac{7m_0^2M^2}{12}+\frac{3m_0^2M'^2}{4}+M^2M'^2-
\frac{m_0^2m_b^2}{2}-\frac{M^2m_bm_d}{2}-M^2m_d^2-M'^2m_d^2\\ &&+\frac{M'^2m_dm_u}{2}-\frac{m_0^2m_u^2}{2}+\frac{m_0^2q^2}{2}-m_d^2q^2+m_b^2m_d^2)\times
e^{-\frac{m_u^2}{M^2}}~e^{-\frac{m_b^2}{M'^2}},
\end{eqnarray*}


\begin{thebibliography}{II}
\bibitem{sumrules}
Coskun Aydin, A.Hakan Yilmaz , Mod.Phys.Lett. A19,2129-2134(2004) ,  V.V.Braguta, A.I.Onishchenko , Phys.Lett. B,591, 267-276 (2004), Takumi Doi, Yoshihiko Kondo , Makoto Oka , Phys.Rept.398,253-279 (2004), R.D. Matheus, F.S. Navarra, M. Nielsen, R. Rodrigues da Silva , arXiv:hep-ph/0310280 .

\bibitem{exotic}
Meng-Lin Du, Wei Chen, Xiao-Lin Chen, Shi-Lin Zhu, Phys. Rev. D 87, 014003 (2013).
\bibitem{FSNavarra}
F. S. Navarra, M. Nielsen, M. E. Bracco, M. Chiapparini, C. L. Schat, Phys. Lett. B 489, 319 (2000).

\bibitem{MNielsen}
F. S. Navarra, M. Nielsen, M. E. Bracco,  Phys. Rev. D 65, 037502 (2002).

\bibitem{MChiapparini}
M.E. Bracco, M. Chiapparini, A. Lozea, F. S. Navarra, M. Nielsen, Phys. Lett. B 521, 1 (2001).

\bibitem{Rodrigues3}
B. O. Rodrigues, M. E. Bracco, M. Nielsen, F. S. Navarra, arXiv:1003.2604[hep-ph].

\bibitem{MEBracco}
M. E. Bracco, M. Chiapparini, F. S. Navarra, M. Nielsen, Phys. Lett. B 659, 559 (2008).

\bibitem{RDMatheus}
R. D. Matheus, F. S. Navarra, M. Nielsen, R. R. da Silva, Phys. Lett. B 541, 265 (2002).

\bibitem{RRdaSilva}
R. R. da Silva, R. D. Matheus, F. S. Navarra, M. Nielsen,  Braz. J. Phys. 34, 236 (2004).

\bibitem{EBracco}
M. E. Bracco, M. Chiapparini, F. S. Navarra, M. Nielsen,  Phys. Lett. B 605, 326 (2005).

\bibitem{ALozea}
M. E. Bracco, A. J. Cerqueira, M. Chiapparini, A. Lozea, M. Nielsen, Phys. Lett. B 641, 286 (2006).

\bibitem{LBHolanda}
L. B. Holanda, R. S. Marques de Carvalho, A. Mihara,  Phys. Lett. B 644, 232 (2007).

\bibitem{Janbazi}
R. Khosravi, M. Janbazi, Phys. Rev. D 87, 016003 (2013).

\bibitem{Song12}
Y. Oh, T. Song and S.H. Lee, Phys. Rev. C 63, 034901 (2001).

\bibitem{123}
Z. Lin and C. M. Ko, Phys. Rev. C 62, 034903 (2000).


\bibitem{Prog12}
B. L. Ioffe, Prog. Part. Nucl. Phys. 56, 232 (2006).

\bibitem{Dosch}
H.G. Dosch, M. Jamin and S. Narison, Phys. Lett. B220, 251 (1989) ; V. M. Belyaev, B. L.
Ioffe, Sov. Phys. JETP, 57, 716 (1982).

\bibitem{Khodjamirian12}
P. Colangelo and A. Khodjamirian, in At the Frontier of
Particle Physics/Handbook of QCD, edited by M. Shifman
(World Scientific, Singapore, 2001), Vol. 3, pp. 1495–1576;
A.V. Radyushkin, in Proceedings of the 13th Annual HUGS
at CEBAF, Hampton, Virginia, 1998, edited by J. L. Goity
(World Scientific, Singapore, 2000), pp. 91–150.

\bibitem{Likhoded}
V.V. Kiselev, A. K. Likhoded, and A. I. Onishchenko,
Nucl. Phys. B569, 473 (2000).

\bibitem{pdg1}
J. Rosner, S. Stone, [Particle Data Group],(URL: http://pdg.lbl.gov).


\bibitem{Wang2}
G. L. Wang , Phys. Lett. B, 633: 492-494(2006).

\bibitem{Bazavov}
A. Bazavov, C. Bernard, C. M. Bouchard , C. DeTar, M. Di Pierro, A. X. El-Khadra, R. T.
Evans, E. D. Freeland, E. Gmiz, Steven Gottlieb, U. M. Heller, J. E. Hetrick, R. Jain, A. S.
Kronfeld, J. Laiho, L. Levkova, P. B. Mackenzie, E. T. Neil, M. B. Oktay, J. N. Simone, R.
Sugar, D. Toussaint, and R. S. Van de Water, Phys. Rev. D 85, 114506 (2012).

\bibitem{Huang2}
Z.G. Wang, T. Huang, Phys. Rev. C 84, 048201 (2011).


\bibitem{Beringer1}
J. Beringer et al., Particle Data Group, Phys. Rev. D 86, 010001 (2012).

\bibitem{Richard}
Z. Guo, S. Narison, J. M. Richard, Q. Zhao, Phys. Rev. D 85, 114007 (2012).

\bibitem{99}
Arxive:1104.2864

\bibitem{Colangelo}
P. Colangelo, F. De Fazio, Eur. Phys. J.C,4, 503-511(1998).

\bibitem{Hungchong}
Hungchong Kim, Su Houng Lee , Eur. Phys. J.C, 22,707-713(2002).

\bibitem{Aliev}
T. M. Aliev, N. K. Pak, M. Savci, Phys. Lett. B, 390:335-340(1997).

\bibitem{Zhu}
Yuan-Ben Dai, Shi-Lin Zhu, Eur. Phys. J.C,6, 307-311(1999).
\end{thebibliography}
\end{document}